\documentclass[uslettersize, 10pt, conference]{ieeeconf}      

\IEEEoverridecommandlockouts                              

\usepackage[OT1]{fontenc} 
\usepackage[english]{babel}
\usepackage{cite}
\usepackage[left=19.1mm,right=19.1mm,top=19.1mm,bottom=19.1mm]{geometry}
\usepackage{amsmath,amssymb,amsfonts}
\usepackage{textcomp}
\usepackage{xcolor}
\usepackage{blindtext}
\usepackage{subfiles} 
\usepackage{mathtools}
\usepackage{adjustbox}
\usepackage{tikz}
\usepackage{svg}
\usepackage{booktabs} 
\usepackage[final]{microtype} 
\usepackage{algorithm} 
\usepackage[algo2e,ruled,vlined]{algorithm2e} 
\usepackage{hyperref}
\hypersetup{hidelinks, breaklinks = true}
\usepackage[capitalise]{cleveref}
\usepackage{mathdots}
\usepackage{yhmath}
\usepackage{cancel}
\usepackage{color}
\usepackage{array}
\usepackage{multirow}
\usepackage{gensymb}
\usepackage{adjustbox}
\usepackage{tabularx}
\usepackage{comment}
\usepackage{extarrows}
\usepackage{booktabs}
\usepackage{diagbox}
\usepackage{graphicx}
\usepackage{subcaption}
\usepackage[normalem]{ulem}
\useunder{\uline}{\ul}{}
\usetikzlibrary{fadings}
\usetikzlibrary{patterns}
\usetikzlibrary{shadows.blur}
\usetikzlibrary{shapes}
\graphicspath{ {./Figures/} }
\usepackage{amsmath}
\usepackage{tikz}
\usepackage{mathdots}
\usepackage{yhmath}
\usepackage{cancel}
\usepackage{color}
\usepackage{array}
\usepackage{multirow}
\usepackage{amssymb}
\usepackage{bm}
\usepackage{gensymb}
\usepackage{tabularx}
\usepackage{extarrows}
\usepackage{booktabs}
\usetikzlibrary{fadings}
\usetikzlibrary{patterns}
\usetikzlibrary{shadows.blur}
\usetikzlibrary{shapes}
\usepackage{algcompatible}
\SetKwComment{Comment}{/* }{ */}

\newtheorem{rem}{Remark}

\DeclareMathOperator*{\argmin}{argmin}

\title{\LARGE \bf
Towards Achieving Cooperation Compliance of Human Drivers\\
in Mixed Traffic

}

\author{Anni Li and Christos G. Cassandras
\thanks{This work was supported in part by NSF under grants CNS-2149511, ECCS-1931600,
DMS-1664644 and CNS-1645681, by ARPAE under grant DE-AR0001282,
and by the MathWorks.}
\thanks{ A. Li, and C. G. Cassandras are
with the Division of Systems Engineering and Center for Information and
Systems Engineering, Boston University, Brookline, MA 02446
(email:\{anlianni; cgc\}@bu.edu).}
}

\begin{document}

\maketitle
\thispagestyle{empty}
\pagestyle{empty}

\begin{abstract}
We consider a mixed-traffic environment in transportation systems, where 
Connected and Automated Vehicles (CAVs) coexist with potentially 
non-cooperative Human-Driven Vehicles (HDVs).
We develop a cooperation compliance control framework to incentivize HDVs to align their behavior with socially optimal objectives using a ``refundable toll'' scheme 
so as to achieve a desired compliance probability for all non-compliant HDVs through a feedback control mechanism combining global with local (individual) components.
We apply this scheme to the lane-changing problem, where a ``Social Planner'' provides references to the HDVs, measures their state errors, and induces cooperation compliance for safe lane-changing through a refundable toll approach.
Simulation results are included to show the effectiveness of our cooperation compliance controller in terms of improved compliance and lane-changing maneuver safety and efficiency when non-cooperative HDVs are present.

\end{abstract}


\section{INTRODUCTION}
Due to rapid urbanization and population growth, the problems of traffic congestion, air pollution, energy consumption, and safety in transportation systems have been well documented \cite{xu2022general,dimitrakopoulos2010intelligent}. The emergence of Connected and Automated Vehicles (CAVs), also known as autonomous vehicles or self-driving cars, has the potential to significantly improve traffic performance and efficiency by better assisting drivers in making decisions so as to reduce accidents, energy consumption, air pollution and traffic congestion \cite{autili2021cooperative,guanetti2018control}. The shared information and cooperation among CAVs facilitate intelligent transportation management through precise trajectory planning, especially in conflict areas such as signal-free intersections, on-ramp merging, roundabouts, and lane-changing scenarios \cite{zhang2023coordinating,xiao2021decentralized,armijos2022cooperative,martin2021traffic,chen2023scalable}. 

While several studies have shown the benefits of CAVs in improving safety and reducing traffic congestion, travel time, and energy consumption, 100\% CAV penetration is not likely to be achieved in the near future. This raises the question of how to coordinate CAVs with Human-Driven Vehicles (HDVs) to enhance traffic efficiency while also ensuring safety, since human driving behavior is stochastic and hard to predict \cite{ghiasi2019mixed,wang2020controllability,malikopoulos2023addressing,xiao2024toward}. To address this challenge, some efforts have concentrated on developing accurate car-following models, as in \cite{zhao2018optimal}, which estimate HDV behavior, as well as on forcing platoons by directly controlling CAVs, as in \cite{mahbub2023safety}. Towards the same goal, learning-based techniques are also used in \cite{guo2020inverse,qu2024model,wang2022federated}. Considering vehicle interactions, game-theoretic approaches are used in \cite{yan2023multi,fu2023cooperative,li2023cooperative} to assist CAVs in evaluating adversarial actions by assuming at least partial knowledge of their opponents. Moreover, since human behavior is uncontrollable and unpredictable, some work \cite{li2024safe,khalil2022connected} focuses on forming CAV coalitions which can ensure safe vehicle interactions despite unknown HDV dynamics. In this case, optimality and performance are traded off for safety with CAVs favoring a conservative operation.

Beyond traditional optimization methods, there has been increasing research focused on integrating social psychology tools and implementing non-monetary management schemes into intelligent transportation systems to improve traffic efficiency. In \cite{alpcan2009nash,barrera2014dynamic,schwarting2019social}, CAVs can make decisions based on quantifying and predicting the social behavior of HDVs and incentivizing them to behave in compliant ways. A concept of social value orientation is defined in \cite{schwarting2019social} to characterize an agent's proclivity for social behavior or individualism and further predict human behavior, while in \cite{shakarami2022steering} the aggregative behavior of non-cooperative price-taking agents is steered towards a desired behavior through a ``nudge'' framework. When solving traffic congestion problems, congestion pricing schemes are proposed in \cite{pigou2017economics,de2011traffic}, where individual drivers have to pay for the negative externalities to ensure traffic efficiency. In \cite{kockelman2005credit}, credit-based congestion pricing (CBCP) is proposed, where drivers receive monetary travel credits to use on roads, while \cite{jalota2023credit} further adopts a mixed economy to let eligible users receive travel credits while ineligible users pay out-of-pocket to use the express lanes. Considering the heterogeneity of human drivers, such congestion pricing schemes can be unfair \cite{brands2020tradable} as they tend to favor wealthier drivers. To overcome these limitations, non-tradable credit-based congestion pricing schemes have been applied as fair and equitable mechanisms in \cite{elokda2022carma}, which proposes a congestion management scheme called ``CARMA" utilizing non-tradable ``karma credits'' for all drivers to bid for the rights to use the express lanes; such credits may be redistributed so that the drivers can choose to benefit themselves now or in the future based on their value of time. Moreover, to deploy social contracts in practice, the use of Distributed Ledger Technologies (DLT) is described in \cite{ferraro2023personalised} to create personalized social nudges and to influence the behavior of users.

In this work, we propose a cooperative compliance framework to incentivize HDVs to cooperate with CAVs, where the control is implemented without any explicit monetary transactions but rather using the following \emph{refundable toll} principle: every user (vehicle) commits a number of tokens to a digital wallet (which can be efficiently implemented through DLT) at the beginning of a driving period. If the user complies with the traffic ``guidance" from a \emph{Social Planner}, then the tokens will be fully refunded. Otherwise, a certain amount of these tokens is removed from the digital wallet. To test the effectiveness of such a cooperation compliance controller, we have implemented this framework in lane-changing problems in a mixed traffic environment to facilitate the coordination between CAVs and HDVs. 
The main contributions of this paper are summarized as follows:
(a) We develop a Cooperation Compliance Control (CCC) framework to incentivize non-compliant HDVs to adjust their behaviors so as to meet system-wide objectives set by a Social Planner.
(b) The refundable toll nature of the proposed scheme is equitable to all system users.
(c) The heterogeneity of HDV behaviors is captured by using predictions based on their own compliance probability. 
(d) We include simulation results of lane-changing maneuvers showing how CCC can improve the performance of the entire traffic network through improved HDV compliance.

The rest of the paper is organized as follows. Section \ref{sec:incentivemechanism} provides the basic structure of the proposed CCC framework, and Section \ref{sec:implementation_in_lanechanging} applies it to the lane-changing problem. Simulation results are provided in Section \ref{sec:simulation}, and we conclude with Section \ref{secV:Conclusions}.

\section{Cooperation Compliance Control Framework}
\label{sec:incentivemechanism}
The goal of the proposed cooperation compliance control scheme is to improve the compliance probability of all users in a transportation network. In particular, we aim to gradually improve the compliance probability $P_i(k)$ for each agent $i\in\{1,2,...,N\}$ over time instants $k=1,2,\ldots$
through a \emph{refundable toll} concept that operates as follows: $(i)$ Every agent/user (driver) owns a digital wallet consisting of virtual tokens that may only be used in the context of the cooperation compliance process.
$(ii)$ An agent commits a number of tokens at the start of a given period, i.e., pays a \emph{toll} in advance.
$(iii)$ The toll is fully refundable as long as the agent complies with every cooperation request received. Otherwise, it is subject to a feedback control mechanism that updates the agent's digital wallet as detailed next.

For each agent $i\in\{1,2,...,N\}$, its compliance behavior over time 
is described by the variable $M_i(k)$, $k=1,2,\ldots$, whose value is selected to be inversely proportional to the deviation of the actual state of agent $i$ from a desired reference set by a \emph{Social Planner} (SP). 
To induce the agents to comply with this reference, we design a \emph{global} cost (penalty) $C(k)$ for all non-compliant agents and a \emph{local} (or individual) cost (penalty) $c_i(k)$ for agent $i$ after non-compliant behavior.
This is inspired by the scheme used in \cite{ferraro2023personalised} based on the following feedback control mechanism:
\begin{align}
\label{eq:global_controller}
&C(k+1) = C(k) + \alpha(Q^*-\frac{1}{N}\sum\limits_{i=1}^{N}M_i(k)),\\
\label{eq:compliance_controller}
&c_i(k+1)=c_i(k)+\beta (Q^*-\Bar{M}_i(k)),
\end{align}
where $Q^*\in[0,1]$ denotes the desired compliance rate (set by the SP), $\alpha,\beta>0$ are tunable control parameters, and $\Bar{M}_i(k)$ denotes the windowed time average of agent $i$'s compliance behavior, given by
\begin{align}
\label{eq:def_aveM}
    \Bar{M}_i(k)=(1-\gamma)\sum_{j=1}^k \gamma^{k-j}M_i(k).
\end{align}
In \eqref{eq:def_aveM}, $\gamma\in(0,1)$ is the (tunable) length of the window over which the average is defined. 
The importance of combining both a global and local cost control is discussed in \cite{ferraro2023personalised}; we will also see its importance in our analysis. 

The compliance probability of agent $i$ after $k$ iterations is assumed to have the following form:
\begin{equation}
\label{eq:compliance_P}
    P_i(k+1) = p(q_i,C(k),c_i(k))
\end{equation}
where $p:\mathbb{R}_+\rightarrow [0,1]$ is a monotonically increasing function in each of its three arguments with $q_i$ indicating the initial proclivity of agent $i$ to comply with the guidance given by the SP. In this paper, we limit ourselves to the following form:
\begin{equation}
\label{eq:compliance_P1}
    P_i(k+1) = p(w_q q_i+w_c C(k) + w_i c_i(k))
\end{equation}
where $w_q,w_c,w_i\in [0,1]$ are the corresponding weights for the initial proclivity, global cost control, and local cost control, respectively, and we adopt the following function:
\begin{equation}
\label{eq:comp_func}
\begin{adjustbox}{width=0.5\linewidth}
    $p(x)=
    \begin{cases}
       1, &\!\! \mathrm{if}\; x>1,\\
       x, &\!\! \mathrm{if}\; 0\leq x \leq 1,\\
       0, &\!\! \mathrm{if}\; x<0.
    \end{cases}$
    \end{adjustbox}
\end{equation}
The precise nature of this function is not critical to our approach, but it obviously affects the overall performance of the control system defined through 
(\ref{eq:global_controller}), (\ref{eq:compliance_controller}), (\ref{eq:compliance_P}). Clearly, there is an opportunity for appropriate Machine Learning (ML) methods to learn its exact form for each agent $i$.

\begin{figure}[hpbt]
    \centering   
    \includegraphics[ width=\linewidth]{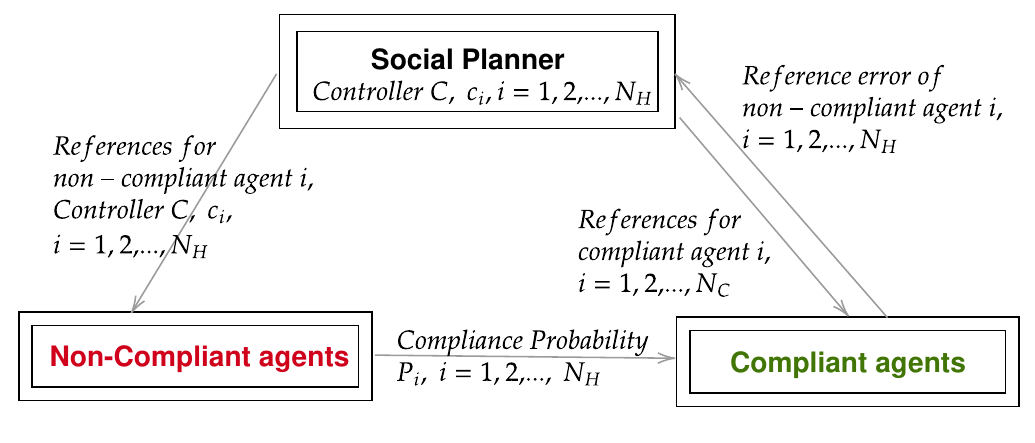} 
    \caption{Cooperation Compliance Control (CCC) framework. Compliant agents are CAVs that provide real-time feedback to the SP. Non-compliant agents are HDVs whose probability of compliance is controlled through refundable tolls.}
    \label{fig:diagram}
\end{figure}

The structure of this control system involving the SP, non-compliant agents, and compliant agents is described in  Fig. \ref{fig:diagram}. Clearly, in this scheme, the ``compliant agents'' are the CAVs (their number is denoted by $N_C$)
which are designed to always follow the guidance prescribed by the SP (i.e., $M_i(k)=1$ for each CAV), while the ``non-compliant agents'' are the HDVs (their number is denoted by $N_H$) whose drivers participate in this system and are subject to the refundable toll concept.

The proposed Cooperation Compliance Control (CCC) framework is general in nature in the sense that it can be applied to any ``rule'' set by the SP at any level of a given transportation system. For example, the SP may evaluate a socially optimal solution to a routing problem and provide reference routes to each agent $i$ (or group of agents according to some criterion, such as enforcing equity, fairness or prioritizing emergency vehicles). In this case, a non-compliant agent is one that violates its prescribed route at any point and incurs some cost that will affect its future behavior. In this paper, we choose to apply CCC to the lane-changing problem where the ``rule'' is for a HDV to cooperate with other agents (either HDVs or CAVs) in order to facilitate a merging maneuver for a vehicle in a slow lane to a fast lane and ultimately improve not only the individual experience of the merging agent (i.e., faster maneuver, lower energy consumption) but also the system-wide performance in terms of throughput and energy efficiency.

\section{Cooperation Compliance Control in the Lane-Changing Problem}
\label{sec:implementation_in_lanechanging}
\begin{figure}[hpbt]
    \centering   
    \includegraphics[ width=\linewidth]{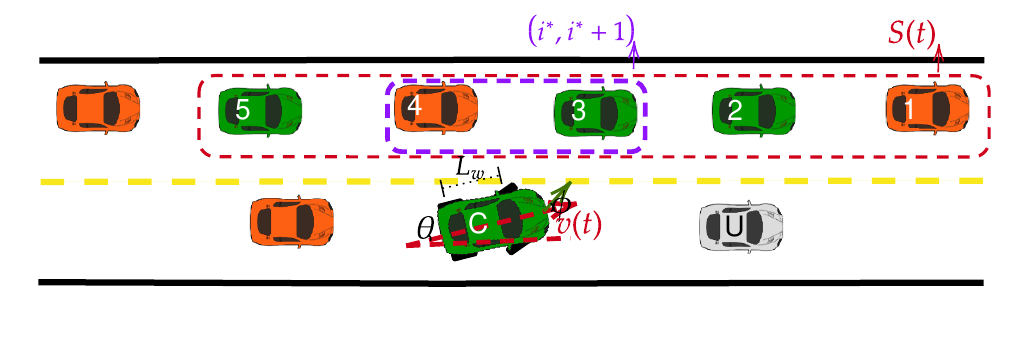} 
    \caption{The basic lane-changing problem. The orange vehicles are HDVs, and the green vehicles are CAVs. The gray vehicle is a slow-moving vehicle.}
    \label{fig:lanechanging}
\end{figure}

In this section, we address the problem of HDV non-compliant behavior in the lane-changing problem, shown in Fig. \ref{fig:lanechanging}, by applying the CCC scheme presented in Sec. \ref{sec:incentivemechanism} and inducing HDVs to cooperate with each other or with CAVs. A key component in this framework is the compliance probability in (\ref{eq:compliance_P}) capturing how human drivers respond  to the social contracts or recommendations issued by the SP, which greatly influences the performance of the traffic network. 
In the lane-changing problem, we decompose the maneuver into a longitudinal component followed by a lateral component. In the longitudinal part, we first compute the optimal maneuver time and trajectories for $C$ to achieve a desired speed that matches the fast lane traffic flow, then solve a decentralized optimal control problem with a fixed terminal time 
for each of the candidate pairs in the fast lane, and pick the one with minimum cost as the optimal merging pair. We assume the SP has perfect knowledge of vehicle states in real-time. This is possible through observations made by the compliant agents (CAVs) with which the SP can exchange information and/or by infrastructure elements such as Road-Side Units (RSUs) which are increasingly present in transportation networks.
The goal of the SP is to minimize maneuver time and energy consumption while alleviating any disruption to the fast lane traffic, thus ultimately providing optimal reference trajectories for all vehicles. 

The basic lane-changing problem is shown in Fig. \ref{fig:lanechanging}, where the green vehicles are CAVs, and the orange vehicles are HDVs. The gray vehicle $U$ is considered as a dynamic obstacle moving at a slower speed, so that a lane-changing maneuver is triggered by CAV $C$ when such an obstacle ahead is detected. Clearly, $C$ has a choice of which vehicles to merge between.
As shown in \cite{armijos2022cooperative}, a key step in this problem is to determine the optimal pair of vehicles in the fast lane that the lane-changing CAV can move in between. 
Typically, the rear vehicle in the merging pair is more critical: a conservative HDV may sacrifice its own performance to decelerate and make space for CAV $C$, while an aggressive uncooperativ HDV may accelerate and force $C$ to abort the maneuver or cause a collision. To achieve the social goal set by the SP and reduce collision risks, the SP issues recommendations or references for HDVs to follow in order to facilitate such a maneuver. However, a human objective may conflict with the social goal and cause disruption and safety issues. Thus, following the CCC framework, we impose cost controls on a non-compliant HDV engaged in a lane-changing maneuver with CAV $C$ based on its deviation from the prescribed behavior so as to improve its compliance probability towards a socially optimal performance as set by the SP.

\textbf{Candidate Vehicle Set.}
In order to determine the optimal merging pair in the fast lane, as shown in Fig. \ref{fig:lanechanging}, we define the set $S(t)=\{1,2,...,M\}$ representing the vehicles in the vicinity of $C$ at time $t$ and containing all candidate pairs when $C$ plans its lane-changing maneuver:
\begin{align}
    S(t) := \{\;i\; |\; x_C(t)-L_r\leq x_i(t)\leq x_U(t)+L_f \}
\end{align}
where $L_r,L_f$ denote the backward and forward distances limited by the CAV's sensors, respectively.
Indices 1 and $M$ correspond to the CAVs farthest ahead and behind $C$, respectively. Details on how this candidate set is determined in practice, can be found in \cite{armijos2022cooperative}. 
The optimal pair in $S(t)$
is labeled $(i^*,i^*+1)$, where $i^*\in\{0,1,...,M\}$. Here, we assume there is a ``virtual" vehicle 0 ahead of 1, with longitudinal position $x_0(t)=+\infty$, and a ``virtual" vehicle $M+1$ following $M$ with $x_{M+1}=-\infty$. If $i^*=0$, it is optimal for $C$ to merge ahead of the whole set $S(t)$ and if $i^*+1=M+1$, it is optimal to merge behind $S(t)$. 
In this problem, the optimal pair $(i^*,i^*+1)$ is defined as the one that corresponds to the minimum cost during $C$'s lane-changing maneuver (details given in the sequel).

\textbf{Vehicle Dynamics.}
For every vehicle in Fig. \ref{fig:lanechanging}, its dynamics are given by a control-affine approximate kinematic bicycle model, as defined in \cite{he2021rule}:
\begin{equation}
\label{eq:CAVs_dynamics}
\begin{adjustbox}{width=0.85\linewidth}
    $\underbrace{\left[\begin{array}{c}
    \dot{x}_i \\
    \dot{y}_i \\
    \dot{\theta}_i \\
     \dot{v}_i   \end{array}\right]}_{\bm{\dot{x}}_i}$
    =
    $\underbrace{\left[\begin{array}{c}
    v_i \cos \theta_i \\
    v_i \sin \theta_i \\
    0 \\
    0    \end{array}\right]}_{f\left(\bm{x}_i(t)\right)}$+
    $\underbrace{\left[\begin{array}{cc}
    0 & -v_i \sin \theta_i \\
    0 & v_i \cos \theta_i \\
    0 & v_i / L_w \\
    1 & 0    \end{array}\right]}_{g\left(\bm{x}_i(t)\right)}
    \underbrace{\left[\begin{array}{l}
    u_i \\
    \phi_i
    \end{array}\right]}_{\bm{u}_i(t)}$
    \end{adjustbox}
\end{equation}
where $x_i(t),y_i(t),\theta_i(t),v_i(t)$ represent the current longitudinal position, lateral position, heading angle, and speed, respectively. $u_i(t)$ and $\phi_i(t)$ are the acceleration and steering angle (controls) of vehicle $i$ at time $t$, respectively. The maneuver starts at time $t_0$ and ends at time $t_f$ when $C$ has completely switched to the target lane. The control input and speed for all vehicles are constrained as follows:
\begin{align}
\label{eq:uv_constraints}
\bm u_{i_{\min}}\leq \bm u_i(t)\leq \bm u_{i_{\max}},~~~v_{i_{\min}}\leq v_i(t)\leq v_{i_{\max}}
\end{align}
where $\bm u_{i_{\min}},\bm u_{i_{\max}}\in\mathbb{R}^2$ denote the minimum and maximum control bounds for vehicle $i$, respectively. $v_{i_{\min}}>0$ and $v_{i_{\max}}>0$ are vehicle $i$'s allowable minimum and maximum speed. Setting $l$ as the width of the lane and the axis $y=0$ as the center line of the slow lane, we have $y_C(t_0)=0$, and the lateral position of all vehicles satisfies $-\frac{l}{2} \leq y_i(t) \leq \frac{3}{2}l$.

\textbf{Safety Constraints.} 
Along the longitudinal direction, we define a speed-dependent safety distance $d_i(v_i(t))$ as the minimum required distance between vehicle $i$ with its current preceding vehicle:
\begin{align}
    d_i(v_i(t))= \varphi v_i(t)+\delta
\end{align}
where $\varphi$ is the reaction time (usually defined as $\varphi=1.8s$ \cite{vogel2003comparison}), $\delta$ is a constant corresponding to the length of the vehicle. After determining the optimal merging pair, i.e., vehicles $i^*$ and $i^*+1$, from the candidate set $S(t)$, we can now define all safety constraints that must be satisfied during $C$'s lane-changing maneuver:
\begin{subequations}
\begin{align}
\label{eq:safety_i_i+1}
    &x_i(t)-x_{i+1}(t)\geq d_{i+1}(v_{i+1}(t)),t\in[t_0,t_f],
    \\
    \nonumber
    &\;\;\;\;\;\;\;\;\;i\in\{1,2,...,M-1\}\\
    \label{eq:safety_CU}
    &x_U(t)-x_C(t)\geq d_C(v_C(t)), t\in[t_0,t_f]\\
    \label{eq:safety_iC}
    &x_{i^*}(t_f)-x_C(t_f)\geq d_C(v_C(t_f))\\
    \label{eq:safety_Ci+1}
    &x_C(t_f)-x_{i^*+1}(t)\geq d_{i^*+1}(v_{i^*+1}(t_f))
\end{align}
\label{eq:safety_equations}
\end{subequations}
The safety between each vehicle pair in the candidate set $S(t)$ in the fast lane is described by constraint \eqref{eq:safety_i_i+1}. \eqref{eq:safety_CU} guarantees the safety between CAV $C$ and vehicle $U$ during the entire maneuver, \eqref{eq:safety_iC} and \eqref{eq:safety_Ci+1} denote the merging safety constraints between $C$ and vehicles $i^*$, $i^*+1$ at the terminal time.

\textbf{Traffic Speed Disruption.} The traffic disruption metric is introduced in \cite{armijos2022cooperative}, which includes both a position and a speed disruption. Each disruption is evaluated relative to its corresponding value under no maneuver. In this paper, we aim to alleviate the fast lane flow disruption during the entire lane-changing maneuver by adopting the speed disruption: 
\begin{align}
    d_v^i(t) = (v_i(t)-v_{d,i})^2
\end{align}
where $v_{d,i}\leq v_{i_{\max}}$ denotes the desired speed of vehicle $i$ which matches the fast lane traffic flow.

\subsection{Longitudinal maneuver}

\textbf{Optimal Control Problem for CAV $C$:}
In the lane-changing problem, as shown in Fig. \ref{fig:lanechanging}, the ``ideal" trajectory for CAV $C$ can be obtained by solving the following optimal control problem (OCP) similar to \cite{armijos2022cooperative}:
\begin{subequations}
    \begin{align}
    \label{eq1:OCP_cavC_cost}   J_C=\min\limits_{t_f,u_C(t)} &\alpha_t(t_f-t_0)+\int_{t_0}^{t_f}\frac{\alpha_{u}}{2}u_C^2(t)dt\\
    \nonumber
    s.t. \; \; &(\ref{eq:CAVs_dynamics}),(\ref{eq:uv_constraints}), \eqref{eq:safety_CU}\\
    \label{eq:C_terminal_v}
    &(v_C(t_f)-v_{d,C})^2\leq \delta_v^2\\
    \label{eq2:OCP_cavC_time}
    &t_0 \leq t_f \leq T
    \end{align}
    \label{eq:OCP_cavC}
\end{subequations}
The cost \eqref{eq1:OCP_cavC_cost} combines the travel time $t_f-t_0$ 
and the energy term $u_C^2(t)$ with the satisfaction of the safe distance constraint \eqref{eq:safety_CU} between vehicles $C$ and $U$ during the entire maneuver. Moreover, the terminal speed $v_C(t_f)$ of CAV $C$ is required to match its desired speed $v_{d,C}$ within tolerance $\delta_v>0$ so as to reduce the speed disruption to the fast lane traffic. Finally, the maneuver time is constrained in \eqref{eq2:OCP_cavC_time} not to exceed a maximum allowable value $T$.

The solution to OCP \eqref{eq:OCP_cavC} provides the ``ideal" maneuver time $t_f^*$, control $u_{C,ref}(t),~t\in[t_0,t_f^*]$, and terminal states $\bm x_{C,ref}(t_f) = [x_C(t_f),y_C(t_f),\theta_C(t_f),v_C(t_f)]^T$, which are considered as the reference for $C$ in the lane-changing maneuver. Based on the ideal terminal states of CAV $C$, the SP can proceed to calculate the optimal trajectories for vehicles $i$ and $i+1$, for each $i\in S(t_0)$ accordingly so as not to disrupt the fast lane traffic flow and minimize their energy consumption. The two OCPs are formulated as follows: 

\textbf{Optimal Control Problem for vehicle $i$:}
\begin{align}
\label{ocp:cav1}
     J_i=\min\limits_{u_{i}(t)} \int_{t_0}^{t_f^*} &\frac{\alpha_{u}}{2}u_{i}^2(t)dt + \alpha_{v} (v_{i}(t_f^*)-v_{d,i})^2 \\
    \nonumber
    s.t. \; \; &(\ref{eq:CAVs_dynamics}),(\ref{eq:uv_constraints}),\eqref{eq:safety_i_i+1},\eqref{eq:safety_iC}
\end{align}

\textbf{Optimal Control Problem for vehicle $i+1$:}
\begin{align}
\label{ocp:hdv}
     J_{i+1}=\min\limits_{u_{i+1}(t)} \int_{t_0}^{t_f^*} &\frac{\alpha_{u}}{2}u_{i+1}^2(t)dt + \alpha_{v} (v_{i+1}(t_f^*)-v_{d,i+1})^2 \\
    \nonumber
    s.t. \; \; &(\ref{eq:CAVs_dynamics}),(\ref{eq:uv_constraints}),\eqref{eq:safety_i_i+1},\eqref{eq:safety_Ci+1}
\end{align}
In \eqref{ocp:cav1} and \eqref{ocp:hdv}, the energy consumption for each vehicle pair ($i$, $i+1$) in the candidate set $S(t_0)$ for $t\in[t_0,t_f^*]$ and the speed disruption at the terminal time $t_f^*$ are minimized, respectively, and the optimal solutions are denoted by $u_{i,ref}(t),~u_{i+1,ref}(t),~t\in[t_0,t_f^*]$. The solution to OCPs \eqref{eq:OCP_cavC}-\eqref{ocp:hdv} can be analytically obtained through standard Hamiltonian analysis similar to OCPs formulated and solved in \cite{armijos2022cooperative}. Thus, we omit the details. 

\textbf{Determine the Optimal Merging Pair:} The optimal merging pair $(i^*,i^*+1)$ is determined by
\begin{align}
\label{opt_pair}
    (i^*,i^*+1) = \argmin\limits_{(i,i+1)\in S(t_0)} J_i+J_{i+1}+J_C
\end{align}
This also provides the optimal solutions $t_f^*$, $\bm x_{C,ref}(t)$, $\bm x_{i^*,ref}(t)$, $\bm x_{i^*+1,ref}(t),$ $t\in[t_0,t_f^*]$ for the optimal merging triplet and the ``ideal" trajectories for the three vehicles to achieve social optimality during the entire lane-changing maneuver. If both vehicles $i^*$ and $i^*+1$ are CAVs, which are controllable, then the lane-changing maneuver can proceed directly under CAV cooperation, as shown in \cite{armijos2022cooperative}. 
However, in a mixed traffic environment, if any one of the vehicles in $(i^*,i^*+1)$ is an HDV, vehicle cooperation and social optimality are no longer guaranteed, and the safety criteria may be violated since human drivers may behave either too conservatively or overly aggressively.

\subsection{Compliance Control from Social Planner}
In order to induce the HDVs to comply with the reference prescribed by the SP so as to achieve the desired social optimality and improve traffic efficiency, we adopt the CCC scheme introduced in Sec. \ref{sec:incentivemechanism} to influence the non-compliant behavior of the HDV.
To address the non-compliant behavior of HDVs in the fast lane, we begin by discretizing time by selecting a fixed length $\Delta$ for each time step such that $t_{k+1}=t_k+k\Delta,k=0,1,2,...$. This allows us to measure the error between references and actual states, and re-solve the OCPs \eqref{eq:OCP_cavC}, \eqref{ocp:cav1}, \eqref{ocp:hdv} to update the references for the merging triplet (the details of this process will be introduced in the following section). Thus, to improve the compliance probability of HDVs towards social optimality, they will incur some (global and local) cost at each sampling time $t_k$. The HDV's compliance probability at the next sampling time $t_{k+1}$ will be updated based on \eqref{eq:compliance_P1}.

\textbf{Evaluate Compliance Behavior:} Let $\bm x_i(t_k)$ be the actual state for vehicle $i$ at time $t_k$ and $\bm x_{i,ref}^k(t),~t\in[t_k,t_f^k]$, $i\in\{i^*,i^*+1\}$ be the reference for vehicle $i$ obtained by solving \eqref{eq:OCP_cavC}, \eqref{ocp:cav1}, \eqref{ocp:hdv}, where $t_f^k$ is the optimal terminal time solved at time $t_k$. Thus, the error between the ideal and real state trajectories (measured by CAVs or the RSUs and reported to the SP) for vehicle $i$ at time $t_k$ 
is given by
\begin{align}
    e_i(t_k) = ||\bm x_i(t_k)-\bm x_{i,ref}^k(t_k)||
\end{align}
Therefore, the compliance behavior of agent $i$ at time $t_k$ is evaluated based on the error $e_i(t_k)$ as follows:
\begin{align}
\label{eq:compliance_behavior}
    M_i(t_k)=\max \{1-\frac{e_i(t_k)}{e_{\max}},0\}
\end{align}
where $e_{\max}$ is a maximum tolerance for the state error specified when we first define the CCC concept. If $e_i(t_k)>e_{\max}$, then $M_i(t_k)=0$ and $i$ is considered as a totally non-compliant agent. The overall compliance behavior of $i$ at $t_k$ is given by $\Bar{M}_i(t_k) \in [0,1]$ in \eqref{eq:def_aveM} by setting $\Bar{M}_i(t_k) \equiv \Bar{M}_i(k)$.
Based on $\Bar{M}_i(t_k)$, we can apply the compliance probability defined in \eqref{eq:compliance_P} to estimate the ensuing behavior of $i$, with the monotonically increasing function $p$ defined in \eqref{eq:comp_func}. 
\begin{rem}
    The definition of $M_i(t_k)$ in (18) exploits the \emph{microscopic} information captured by the vehicle states, hence $e_i(t_k)$. A simplified \emph{macroscopic} alternative is to define $M_i(k) \in \{0,1\}$ measuring whether $i$ complies with a request or not at any request time $t_k$ (e.g., $M_i(k)=0$ if HDV $i$ does not allow $C$ to merge at $t_k$). The CCC scheme still applies at this level, entailing minimal real-time information exchanges.
\end{rem}

\textbf{Estimate Fast Lane HDV Behavior:} 
For the HDV involved in the maneuver, we estimate its trajectory by assuming that the human driver considers the following factors: (i) minimizing its acceleration/deceleration when changing speeds so as to save fuel, (ii) maintaining a constant speed or reducing the deviation with some desired speed $v_{d,H}$, (iii) keeping a safe distance with its preceding vehicle. However, the trajectories that meet the above goals may conflict with the references, and the deviated behavior will be penalized by the SP. The human driver must take the penalties into account, which leads to another factor: (iv) minimizing the cost imposed by the SP.

At time $t_k$, we formulate an optimal control problem for HDV $i$ to derive actual trajectories after including all the factors above:
\begin{align}
    \nonumber
        &\min_{u_i(t)} \int_{t_k}^{t_{f}^{k}} P_i(t_k)(u_i(t)-u_{i,ref}^k(t))^2 \\
        \label{eq:realhdv_cost}
        &\;\;\;\;\;\;\;\;\;\;+ (1-P_i(t_k))((u_i(t))^2+(v_i(t_f^k)-v_{i,d})^2) dt, \\
        \nonumber
        &\;\;s.t. \;\; \eqref{eq:CAVs_dynamics}, \;\eqref{eq:uv_constraints},\; \eqref{eq:safety_i_i+1}
 \end{align}
The cost in \eqref{eq:realhdv_cost} consists of two parts: (i) the \emph{social} goal to minimize the deviation of the real HDV control $u_i(t),~t\in[t_k,t_f^k]$ from the reference $u^k_{i,ref}(t)$; (ii) the \emph{selfish} goal to minimize its own energy consumption and speed disruption. The presence of the compliance probability $P_i(t_k)$ 
emphasizes the effect of HDV compliance to its reference at $t_k$. The solution $u_i(t),t\in[t_k,t_{f}^{k}]$ to \eqref{eq:realhdv_cost} is the actual trajectory of HDV $i$ when its compliance probability is $P_i(t_k)$. 

As for CAVs, since they are assumed to be controllable and cooperative, their compliance probability is set as $P_i(t)=1$ for any CAV $i$ in the fast lane at any time $t$ during the maneuver. Hence, CAVs will automatically follow the guidance from the SP and the actual CAV trajectories can be similarly obtained by solving OCP \eqref{eq:realhdv_cost} with $P_i(t_k)=1$. 

At the next sampling time $t_{k+1}$, the SP will use the error $e_i(t_{k+1})$ between $\bm x_i(t_{k+1})$ and $\bm x_{i,ref}(t_{k+1})$ (recall that the SP obtains state measurements from CAVs or RSUs), update $M_i(t_{k+1})$ defined in \eqref{eq:compliance_behavior}, adjust it through the compliance controllers $C(k+1),~c_i(k+1)$ in \eqref{eq:global_controller} and \eqref{eq:compliance_controller}, and improve $i$'s compliance probability for $t>t_{k+1}$.

\subsection{Lateral Maneuver}
The lateral maneuver is initiated at time $t^l$ when CAV $C$ is longitudinally safe with its preceding vehicle $U$ and the two vehicles $(i^*,i^*+1)$ in the optimal merging pair. The value of $t^l$ is determined as:
\begin{align}
\nonumber
    t^l = \min\{t_k: &(x_U(t_k)-x_C(t_k)\geq  d_C(v_C(t_k))) \wedge \\
    \nonumber
    &(x_{i^*}(t_k)-x_C(t_k)\geq  d_C(v_C(t_k))) \wedge\\
    \nonumber
    &(x_C(t_k)-x_{i^*+1}(t_k)\geq  d_{i^*+1}(v_{i^*+1}(t_k))),\\
    \label{eq:long_safe}
    &k=0,1,2,...\}
\end{align}

At the starting time $t^l$, with initial states $[x_C(t^l),0,0,v_C(t^l)]$ of CAV $C$ and the terminal lateral position $y_C(t_f^l)=l$, we can formulate an optimal control problem for its lateral maneuver:
\begin{align}
    \min_{\phi_C(t),t_f^l} &\int_{t^l}^{t_f^l} \frac{1}{2}\alpha_{\phi}\phi_C^2(t)dt + \alpha_t(t_f^l-t^l)\\
    \nonumber
    s.t.&  ~\eqref{eq:CAVs_dynamics},~\eqref{eq:uv_constraints},~-\frac{l}{2} \leq y_i(t) \leq \frac{3}{2}l
\end{align}
where the cost function combines both the lateral maneuver time and energy consumption (approximated through the integral of $\phi_C^2(t)$) with weights $\alpha_t,\alpha_{\phi}$, respectively. We assume the speed $v_C(t)$ is a constant during $[t^l,t_f^l]$. Details of the solution to this problem are given in \cite{armijos2022cooperative}, where it is also shown that the lateral phase is much shorter compared to the longitudinal maneuver's duration.

\subsection{Lane-Changing Process}
In this section, we describe the process of applying the CCC for HDVs in the lane-changing maneuver performed by CAV $C$ in Fig. \ref{fig:lanechanging}. Assuming CAV $C$ starts to change lanes at time $t_0$, we proceed as follows:

\emph{1) Step 1:} The SP measures the initial states $\bm{x}_i(t_0)=[x_i(t_0),y_i(t_0),\theta_i(t_0),v_i(t_0)]^T$, $i\in\{C,1,2,...,M\}$ for all vehicles, finds the candidate set $S(t_0)$, and predetermines the initial compliance probability for all HDVs, the constants $\alpha,\beta$ in controllers $\eqref{eq:global_controller}$ and \eqref{eq:compliance_controller}. The weights $\alpha_t,\alpha_u,\alpha_v$, desired speed $v_{i,d}$ in OCPs \eqref{eq:OCP_cavC}-\eqref{ocp:hdv} are given by each vehicle.

\emph{2) Step 2:} The SP solves \eqref{eq:OCP_cavC}-\eqref{ocp:hdv} for all candidate pairs $(i,i+1),~i\in S(t_0)$, applying \eqref{opt_pair} to get the optimal merging pair $(i^*,i^*+1)$ for CAV $C$, and the reference for vehicles $(i^*,i^*+1)$. Then each vehicle solves \eqref{eq:realhdv_cost} based on its own compliance probability and obtains its actual trajectory: $\bm x_i(t),~t\in[t_0,t_f]$.

\emph{3) Step 3:} At each sampling time $t_k=t_0+k\Delta,k=1,2,...$, where $t_k<t_f$, the SP obtains the states $\bm{x}_i(t_k)=[x_i(t_k),y_i(t_k),\theta_i(t_k),v_i(t_k)]^T$, $i\in\{C,1,2,...,M\}$ for all vehicles, calculates the corresponding error $e_i(t_k)$ to get $\Bar{M}_i(t_k)$ for each vehicle $i$, and updates $\eqref{eq:global_controller}$, \eqref{eq:compliance_controller}, and  \eqref{eq:compliance_P}.

\emph{4) Step 4:} The SP re-solves the OCPs \eqref{eq:OCP_cavC}-\eqref{ocp:hdv} to get the optimal merging pair $(i^*,i^*+1)$ for all $i\in S(t_k)$. With the updated compliance probability, vehicles solve \eqref{eq:realhdv_cost} to get their actual trajectories $\bm x_i(t),~t\in[t_k,t_f^k]$.

\emph{5) Step 5:} CAV $C$ checks if the longitudinal safety constraint \eqref{eq:long_safe} is satisfied to start the lateral part. If \eqref{eq:long_safe} is satisfied, then we proceed to execute the lateral maneuver. Otherwise, go back to \emph{Step 3} at the next sampling time $t_{k+1}$.

\section{Simulation Results}
\label{sec:simulation}
In this section, we test the effectiveness of the proposed CCC framework in the lane-changing problem using a simulation setting as in Fig. \ref{fig:lanechanging}, in which the candidate set $S(t_0)$ contains vehicles 1 to 5 in the fast lane. Vehicles 1 and 4 are HDVs, while vehicles 2,3,5 are CAVs. The allowable speed range is $v\in[15,35] m/s$, and the vehicle acceleration is limited to $u\in[-7,3.3] m/s^2$. The desired speed for the CAVs is considered as the traffic flow speed, which is set to $30 m/s$. The desired speed for the HDV is assumed to be the same as its initial speed. To guarantee safety, the inter-vehicle safe distance is given by $\delta = 1.5 m$, the reaction time is $\varphi = 0.6 s$. In the penalty $\eqref{eq:compliance_controller}$, set the constant $\beta=0.1$, desired compliance probability $Q^*=1$, and the constant $\gamma=0.7$ in \eqref{eq:def_aveM} when computing $\Bar{M}_i$. In the compliance probability \eqref{eq:compliance_P}, the weights $w_q=1, w_c = 0.5, w_i=0.5$.
We obtained numerical solutions for all optimization problems using an interior point optimizer (IPOPT) on an Intel(R) Core(TM) i7-8700 3.20GHz. 

\begin{figure}[hpbt]
    \centering   
    \includegraphics[width=0.65\linewidth]{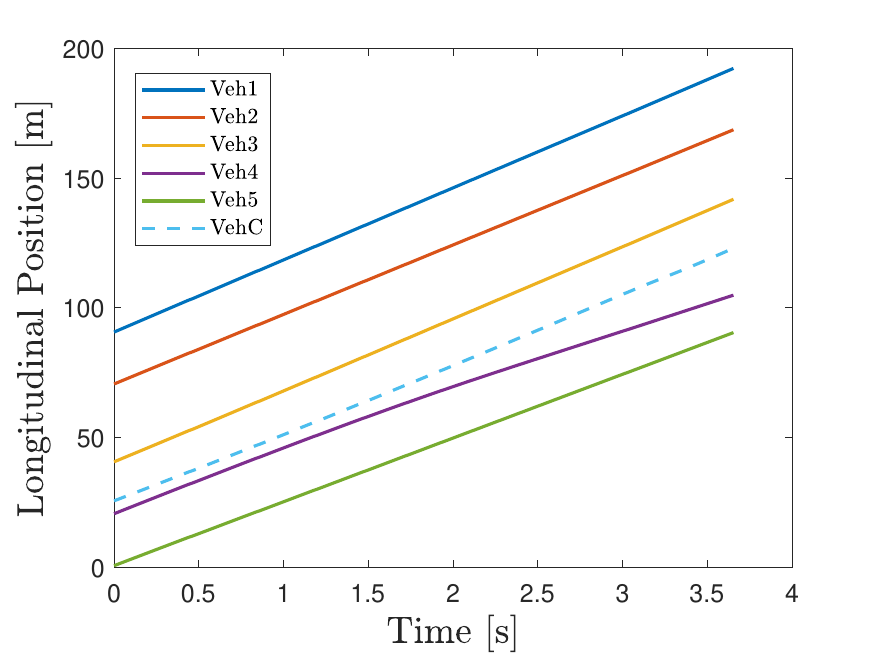} 
    \caption{The longitudinal position of CAV $C$ and all vehicles $i\in S(t_0)=\{1,2,3,4,5\}$ during the entire maneuver.}
    \label{fig:x-t}
\end{figure}

\begin{figure*}[t]
	\centering	
         \begin{subfigure}{0.28\linewidth}
            \centering 
        \includegraphics[width=\linewidth]{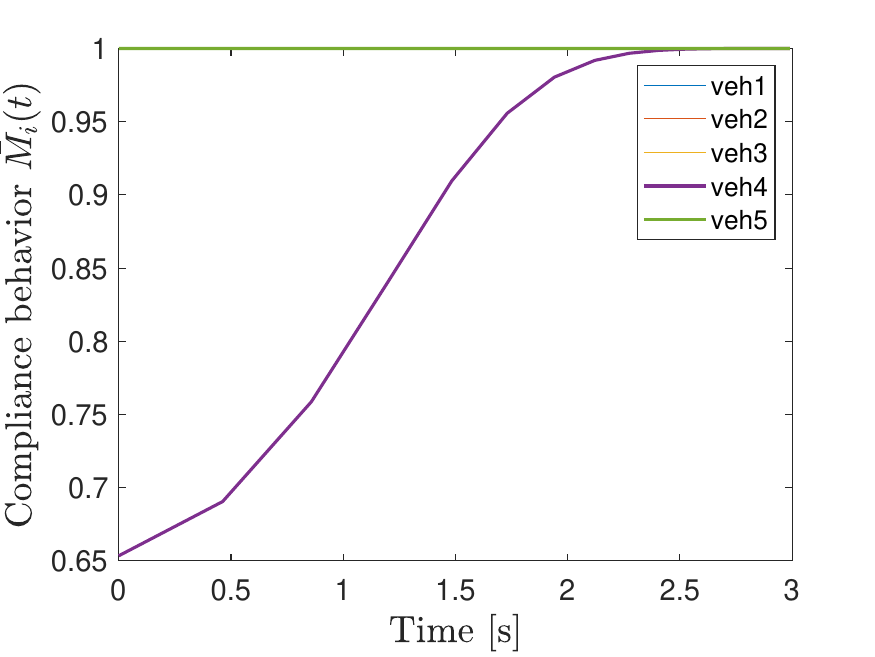}  
        \caption{\centering{Compliance behavior $\Bar{M}_i(t)$.}}
        \label{fig:comp_prob_behavior}
        \end{subfigure} 
	\begin{subfigure}{0.28\linewidth}
            \centering 
        \includegraphics[width=\linewidth]{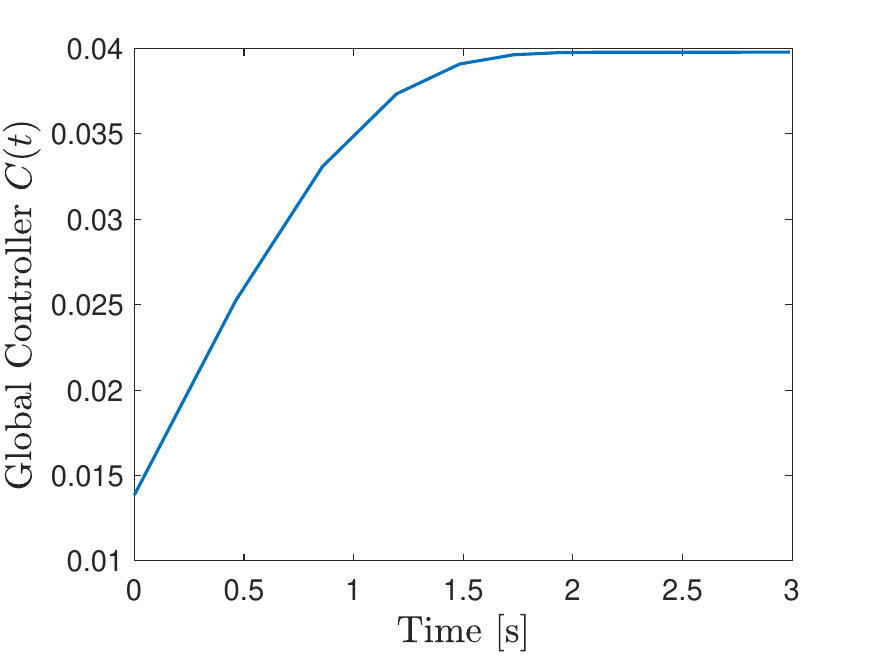}  
        \caption{\centering{Global Controller $C(t)$}}
        \label{fig:C}
        \end{subfigure} 
        \begin{subfigure}{0.28\linewidth}
            \centering 
        \includegraphics[width=\linewidth]{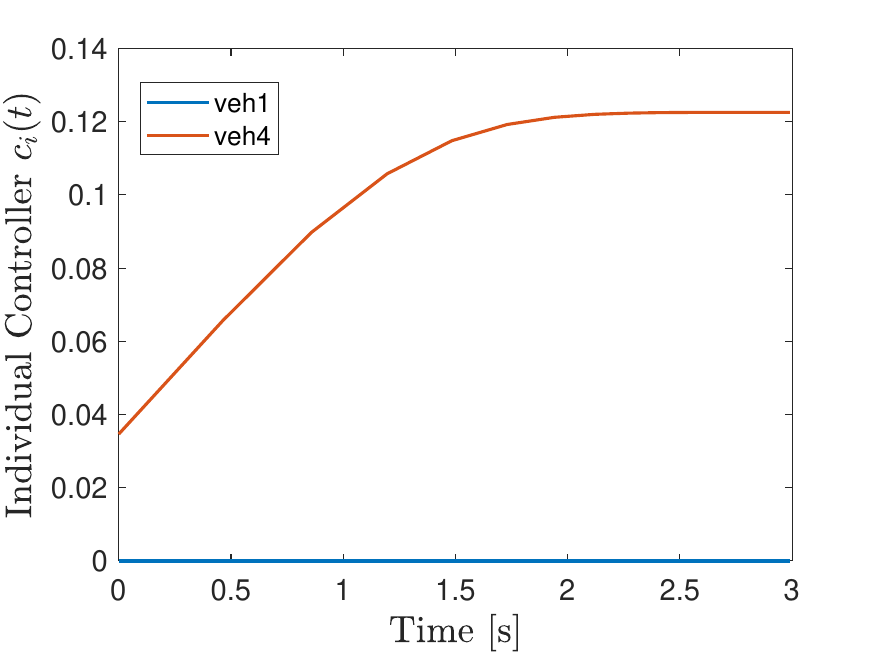}  
        \caption{\centering{Individual Controller $c_i(t)$}}
        \label{fig:ci}
        \end{subfigure} 	
	\caption{The changing of compliance behavior, global controller (penalty), and individual controller for the two HDVs 1 and 4 in the fast lane during the entire lane-changing maneuver for CAV $C$.}
 \label{fig:compliance_multiple}  
	\centering
\end{figure*}

We set the width of the lane as $l=4m$ and the initial states for all vehicles in $S(t_0)$ as
$\bm x_1(t_0)=[90m,4m,0rad,28m/s]$, $\bm x_2(t_0)=[70m,4m,0rad,27m/s]$, $\bm x_3(t_0)=[40m,4m,0rad,27m/s]\\$, $\bm x_4(t_0)=[20m,4m,0rad,26m/s]$, $\bm x_5(t_0)=[0m,4m,0rad,25m/s]$. The parameters are set as $\alpha_u=0.4,\alpha_t=0.6$ to balance the energy consumption and travel time in the optimal control problem for CAV $C$. The initial compliance proclivities are set as $q_1=0,~q_4=0.3,$ and $q_2=q_3=q_5=1$ for all 5 vehicles in $S(t_0)$. After applying the cooperation compliance framework to influence the non-compliant behaviors of HDV 1 and 4, the longitudinal positions of vehicle $i,i\in S(t_0)=\{1,2,3,4,5,C\}$ are summarized in Fig. \ref{fig:x-t}. Under the above initial conditions, the optimal option for $C$ is to merge between CAV 3 and HDV 4. As shown in Fig. \ref{fig:x-t}, HDV 4 (purple curve) decelerates to make enough space for $C$ to merge ahead of it so as to avoid a possible collision. The merging safety constraints along the longitudinal direction between $C$, 3 and 4 are satisfied at time $t^l=3s$, which implies safety for $C$ to start the lateral maneuver.

\textbf{Computational Cost:}
Our results took an average of 200 $ms$ to solve the OCP \eqref{eq:OCP_cavC} for CAV $C$, and took an average of 30 $ms$ to solve each of OCP \eqref{ocp:cav1} and \eqref{ocp:hdv} for fast lane vehicles. 

\begin{figure*}[t]
	\centering	
         \begin{subfigure}{0.28\linewidth}
            \centering 
        \includegraphics[width=\linewidth]{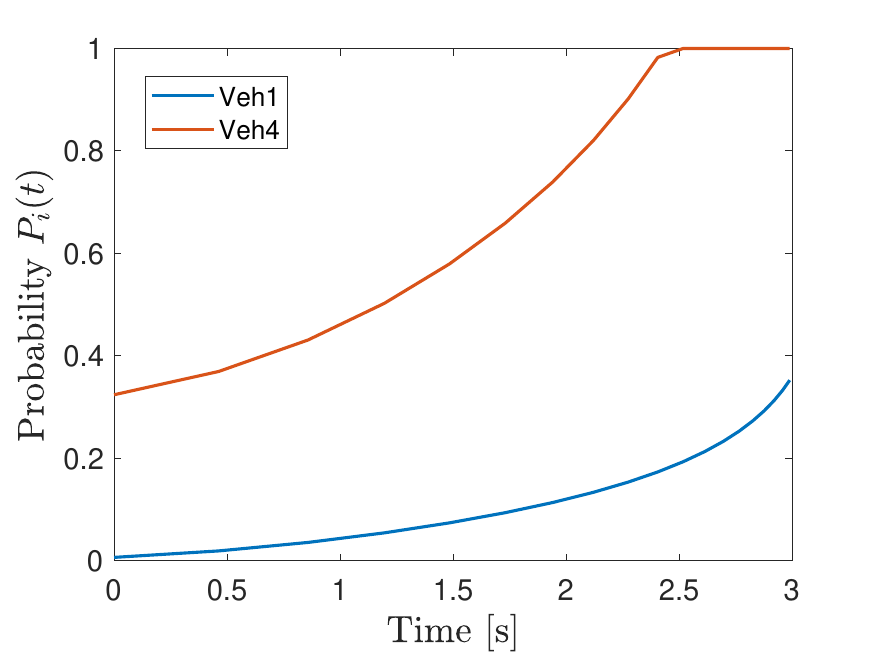}  
        \caption{\centering{Under both global controller $C(t)$ and local controller $c_i(t)$}}
        \label{fig:p-t-C-ci}
        \end{subfigure} 
	\begin{subfigure}{0.28\linewidth}
            \centering 
        \includegraphics[width=\linewidth]{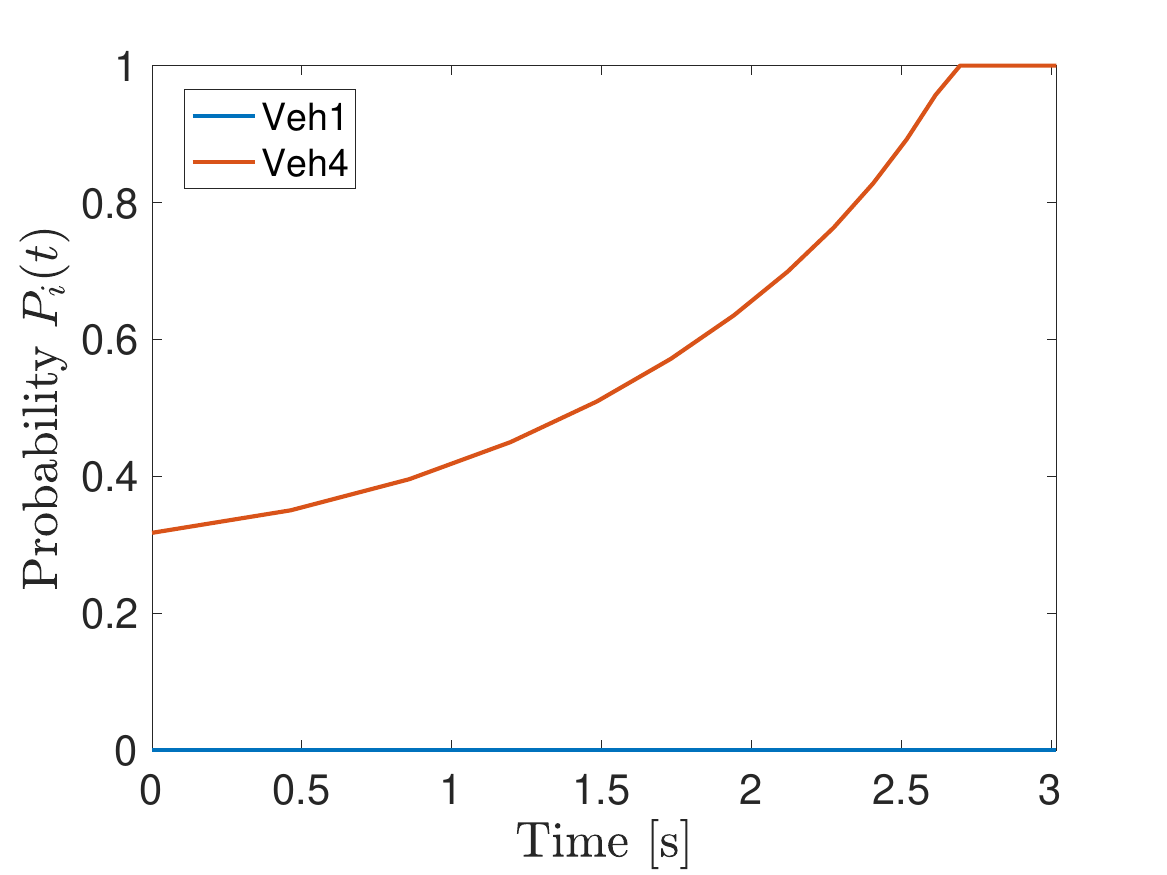}  
        \caption{\centering{Under local controller $c_i(t)$ only}}
        \label{fig:p-t-ci}
        \end{subfigure} 
        \begin{subfigure}{0.28\linewidth}
            \centering 
        \includegraphics[width=\linewidth]{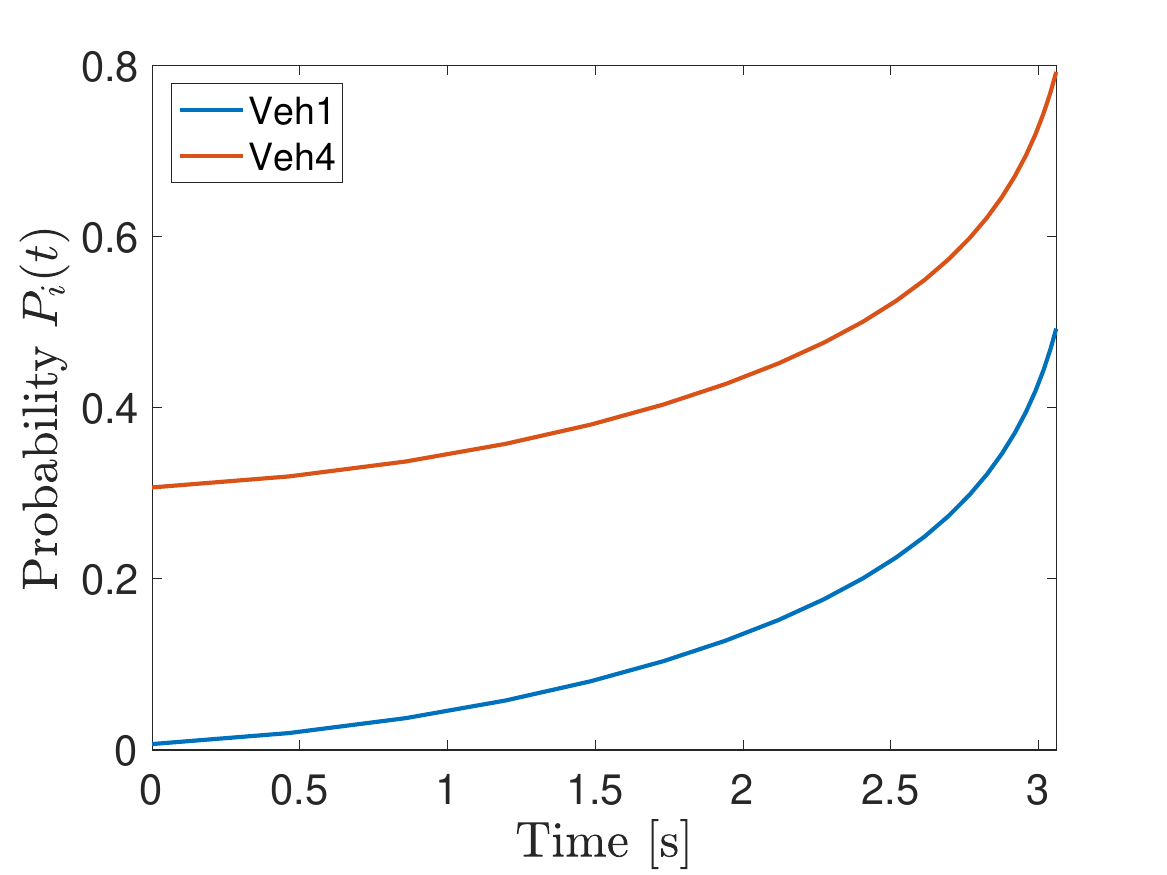}  
        \caption{\centering{Under global controller $C(t)$ only}}
        \label{fig:p-t-C}
        \end{subfigure} 	
	\caption{The change of the compliance probability for HDVs 1 and 4 in a lane-changing maneuver.}
 \label{fig:p-t}  
	\centering
\end{figure*}

The compliance behaviors $\Bar{M}_i(t)$ for all vehicles $i\in S(t_0)$ and cost control evolutions $C(t),~c_i(t)$ for the two HDVs 1 and 4 are summarized in Fig. \ref{fig:compliance_multiple}. 
In Fig. \ref{fig:compliance_multiple}(a), HDV 1 and all CAVs 2, 3, 5 in the candidate set comply with the reference (the four curves overlap and are equal to 1) except for HDV 4. The reason is that HDV 1 is the leading vehicle and its selfish goal is consistent with the social optimal objective, while CAVs are all cooperative and controllable so that their compliance probabilities are equal to 1. Since HDV 4 is the rear vehicle in the optimal pair, its selfish goal conflicts with the social goal because it has to sacrifice its own performance to make enough space for $C$ to merge ahead. As seen in the purple curve, after being penalized by the two controllers, the compliance behavior of HDV 4 eventually increases to 1.
Figures \ref{fig:compliance_multiple}(b) and \ref{fig:compliance_multiple}(c) show the evolutions of the global and local controllers $C(t),~c_i(t)$, respectively. The global controller $C(t)$ for all HDVs converges to 0.04, and local controller $c_i(t)$ for HDV 1 and 4 converges to 0 and 0.12, respectively, indicating that a non-compliant HDV will ultimately comply with the prescribed reference provided by the SP,
confirming the effectiveness of the CCC framework and demonstrating how agents are not penalized with unbounded costs. In summary, Fig. \ref{fig:compliance_multiple} illustrates how after incurring a cost for non-compliant behavior at each sampling time $t_k$, a non-compliant agent will behave increasingly more cooperatively and improve its compliance behavior in subsequent maneuvers. 

\begin{table*}[]
\centering
\caption{\fontsize{10}{\baselineskip} Travel time and energy consumption for the lane-changing triplet under different initial compliance probabilities.}
\label{tab:cost_comparison}
\begin{tabular}{ccccccccc}
\toprule
\multirow{2}{*}{\textbf{Initial   Compliance Probability}} & \multicolumn{4}{c}{\textbf{Maneuver Time} {[}s{]}} & \multicolumn{4}{c}{\textbf{Energy Consumption}} \\ \cmidrule{2-9} 
                                                  & 0        & 0.2      & 0.5      & 0.8      & 0        & 0.2     & 0.5     & 0.8     \\ 

\midrule
\textbf{Compliance Control}                                        & 3.7      & 3.68     & 3.63     & 3.51     & 214.47   & 130.7   & 88.34   & 78.85   \\
\textbf{Baseline}                                          & Inf      & Inf      & 3.65     & 3.51     & Inf      & Inf     & 90.86   & 80.08   \\ \bottomrule
\end{tabular}
\end{table*}

The significance of using both a global and local controller in \eqref{eq:global_controller} and \eqref{eq:compliance_controller} for the two HDVs 1 and 4 is captured in Fig. \ref{fig:p-t}(a): both are increasing with the increasing rate for HDV 4 being faster than HDV 1. The reason is that HDV 4 incurs both global and local costs for its non-compliant behavior, while HDV 1's trajectories, in this case, comply with the reference so that its increase comes only from the global penalty. Moreover, to highlight the need to use both global and local controllers, we provide the variation of compliance probability $P_i(t)$ under the local controller $c_i(t)$ only and under the global controller $C(t)$ only for both HDVs 1 and 4 in Fig. \ref{fig:p-t}(b) and Fig. \ref{fig:p-t}(c), respectively. As shown in Fig. \ref{fig:p-t}(b), if only the local controller $c_i(t)$ is applied, HDV 1's compliance probability does not increase since it already complies with the reference in this case. This deviates from our goal of nudging all non-compliant vehicles to improve their compliance rate after each maneuver, so that the entire network can benefit from the CCC framework in the future. Moreover, the convergence rate of HDV 4 without the global controller $C(t)$ is slower than the one in Fig. \ref{fig:p-t}(a). If only the global cost $C(t)$ is applied, as shown in Fig. \ref{fig:p-t}(c), HDVs 1 and 4 have the same increasing rate and HDV 4 fails to achieve the desired compliance probability by the end of the maneuver, while HDV 1 incurs a higher cost (compared to Fig. \ref{fig:p-t}(a)) even though it complies with the reference in the lane-changing maneuver. 

In summary, the local controllers ensure \emph{fairness} in inducing higher costs for less compliant agents, while at the same time the global controller ensures that all agents contribute towards the desired socially optimal goal.


Therefore, at the end of a single lane-changing maneuver, the compliance probability for all HDVs will be updated based on the CCC framework, and the efficiency of the entire traffic network will be improved at subsequent time instants with new lane-changing maneuvers. To test the effectiveness of the cooperative compliance framework in improving traffic efficiency, we have collected the travel time and energy consumption for the lane-changing triplet under different initial compliance probabilities, and the simulation results are summarized in Tab. \ref{tab:cost_comparison}. The initial compliance probability pertains to HDV 4, while ``Compliance Control" corresponds to the results after applying the cooperation compliance controllers $C,~c_i$, and ``Baseline" corresponds to the results without applying any CCC. Under the same initial conditions, after applying the compliance controller, when the initial compliance probability of HDV 4 increases from 0 to 0.8, the maneuver time decreased by 6.4\% from 3.75s to 3.51s, and the energy consumption for the triplet decreased by 75.8\% from 214.47 to 78.85.
It is worthwhile observing that without compliance control in the baseline case, the lane-changing maneuver is infeasible (denoted by Inf) when the initial compliance probability for HDV 4 is less than 0.2. In addition, under the same initial compliance probability of HDV 4, the total cost of maneuver time and energy consumption in the baseline case is larger than in the compliance control results.

\section{CONCLUSIONS AND FUTURE WORK}
\label{secV:Conclusions}
We have proposed a cooperation compliance control (CCC) framework to incentivize non-compliant HDVs to align their behavior with socially optimal objectives using a ``refundable toll'' scheme with a feedback control mechanism combining a global and local (individual) components. We have applied this scheme to lane-changing problems with promising results, understanding that there remains much to be done in terms of applying CCC to a variety of compliance problems at all levels of transportation networks with mixed traffic, learning human reactions from past data, and exploring how to deal with ``stubborn" agents whose behavior is hard to affect.

\bibliographystyle{IEEEtran}
\bibliography{cmp}

\end{document}